# Magnetic field induced emergent inhomogeneity in a superconducting film with weak and homogeneous disorder

Rini Ganguly[a], Indranil Roy[a*], Anurag Banerjee[b], Harkirat Singh[a], Amit Ghosal[b†] and Pratap Raychaudhuri[a‡]

[a]Tata Institute of Fundamental Research, Homi Bhabha Road, Colaba, Mumbai 400005, India.

[b] Indian Institute of Science Education and Research Kolkata, Mohanpur Campus, Nadia 741252, India.

When a magnetic field is applied, the mixed state of a conventional Type II superconductor gets destroyed at the upper critical field $H_{c2}$, where the normal vortex cores overlap with each other. Here, we show that in the presence of weak and homogeneous disorder the destruction of superconductivity with field follows a different route. Starting with a weakly disordered NbN thin film ( $T_c \sim 9K$ ), we show that under the application of magnetic field the superconducting state becomes increasingly granular, where lines of vortices separate the superconducting islands. Consequently, phase fluctuations between these islands give rise to a field induced pseudogap state, which has a gap in the electronic density of states but where the global zero resistance state is destroyed.

[*] E-mail: indranil.roy@tifr.res.in
[†] E-mail: ghosal@iiserkol.ac.in
[‡] E-mail: pratap@tifr.res.in

I. Introduction

Over the past decade, the notion of emergent granularity has evolved from a theoretical proposition[1,2,3] to a new paradigm to understand the superconducting state in presence of strong homogeneous disorder[4,5,6,7,8,9,10]. Clean s-wave superconductors which are well described by Bardeen-Cooper-Schrieffer (BCS) theory[11] are characterized by an energy gap ($\Delta$) in the electronic density of states centered around the Fermi level, and two sharp coherence peaks at the gap edge. Theoretically, $\Delta$ is the Cooper pairing energy scale which determines the superconducting transition temperature, $T_c$, whereas the coherence peaks signify the establishment of long range phase coherence. Since the late fifties[12,13], it was known that $\Delta$ remains finite even at strong (non-magnetic) disorder, leading to the belief that the superconducting transition would also remain robust against disorder. However, this conclusion is invalidated by the emergence of granularity. It is now understood that in the presence of strong disorder, the superconducting state can segregate into superconducting and insulating regions, where zero resistance is achieved through Josephson tunneling between superconducting islands. Consequently, the superconducting state can get destroyed through phase fluctuations between the islands, even when the pairing amplitude remains finite[1,14]. Experimentally, this manifests as a pseudogap[15,5,6,16], which persists well above $T_c$, where zero resistance state is destroyed. Experimental and theoretical evidences also suggest that at a critical disorder Cooper pairs can eventually get localized, giving rise to an insulator made out of Cooper pairs[17,18].

Another aspect of strongly disordered superconductors, namely, the magnetic field induced superconductor-insulator transition[19,20] has also attracted considerable attention. Under the application of magnetic field, the ground state of several strongly disordered superconductors transform into an insulator, with resistance exceeding $10^9$ Ohm. While no consensus has so far emerged on the origin of this insulator state, it is now widely accepted that the insulating state is related in some way to the superconducting correlations[21,22,23]. Several theoretical scenarios, such as Coulomb blockade in a granular superconducting state[24], Boson localization[25], as well as theories invoking charge vortex duality[26,27] have been proposed to explain this phenomenon. From an experimental standpoint it is therefore important to obtain microscopic information on the evolution of the superconducting state with magnetic field in order to discriminate between various possibilities.

Here, using low temperature scanning tunneling spectroscopy (STS) we investigate the magnetic field evolution of the superconducting state in a weakly disordered NbN thin film[28]. The sample under investigation is an NbN film with $T_c \sim 9$ K, corresponding to $k_F l \sim 4$ (where $k_F$ is the Fermi wave vector and $l$ is the electronic mean free path) [5]. In the present context, the word "weak disorder" is used to delineate the regime where the superconducting energy gap in zero field vanishes at the same temperature where resistance appears as opposed to the strong disorder regime where a pseudogap state appears[29]. The normal state exhibits a weak negative temperature coefficient of resistance with $\frac{R(300K)}{R(15K)} \sim 0.78$. The central result of this paper is that when a magnetic field ($H$) is applied on the sample, the superconducting state becomes inhomogeneous, in a manner similar to what disorder alone would have done at much larger strength. This experimental observation is backed by numerical simulations which show that the flux tubes enter a disordered superconductor at locations where disorder partially suppresses the superconducting correlations. Riding on the top of this inhomogeneous background a vortex line wipes out the pairing amplitude in a region of linear size of about the coherence length, $\xi$, by accommodating the phase-twist around itself. The resulting superconducting state contains regions, few tens of nanometers in size, where the superconducting order parameter is finite, separated by lines of vortices where the superconducting order is suppressed. Consequently, the system exhibits a field induced pseudogap that progressively widens as the magnetic field is increased.

**II. Sample details and experimental methods**

*Sample.* The epitaxial NbN thin film with thickness ~25 nm was deposited on single crystalline MgO (100) substrate using reactive pulsed laser deposition using a KrF excimer laser ( 248 nm ). A pure Nb target was ablated in an ambient $N_2$ atmosphere of 40 mTorr, while the substrate was kept at $600^0$ C. For the ablation process, laser pulses with energy density 200 mJ/mm$^2$ were used with a repetition rate of 10 Hz. To maintain a pristine surface while transporting the sample for STS measurements, the sample was transferred *in-situ* in an ultra-high vacuum suitcase with base pressure ~ $10^{-10}$ Torr and transferred to the low-temperature scanning tunneling microscope (STM) without exposure to air. Transport (and magnetic) measurements were performed on the sample after all STS measurements were completed. The resistance measurements as a function of temperature and magnetic field were performed using conventional 4-probe technique, using a current of 0.5 mA.

*Scanning tunneling spectroscopy measurements.* All STS measurements were performed in a home built STM[30] operating down to 350 mK and fitted with a superconducting solenoid with maximum field of 90 kOe. The STM tip was made out of a mechanically cut Pt-Ir tip, which was sharpened *in-situ* by field emission on an Ag single crystal. The tunneling conductance $\left(G(V) = \frac{dI}{dV}\bigg|_V\right)$ was measured by adding a 150 µV, 2 kHz ac voltage to the d.c. bias voltage (*V*) and recording the a.c. response in the tunneling current (*I*) using standard lock-in technique. For the conductance maps at fixed bias voltage, the a.c. response in the tunneling current was measured while the tip is rastered over the sample surface. The full area spectroscopy was performed by stabilizing the tip at every point, then momentarily switching off the feedback loop and sweeping V from +5mV to -5mV and recording the a.c. response in the tunneling current as a function of bias voltage. Transport measurements were carried out on the same sample after completing all STS measurements.

## III. Results

*The zero field superconducting state*

The tunneling conductance between a normal tip and a superconductor, measured using an STM provides the most direct access to the local density of states in a superconductor. We characterize the zero field state by measuring the tunneling conductance spectra (*G(V)* vs. *V*) over 200 nm × 200 nm area at 450 mK. The tunneling conductance reveals a uniform energy gap, and the presence of coherence peak at the gap edge over the entire area (Fig **1a**). To look at the spatial variation of the zero bias conductance (ZBC) and the coherence peak height we normalize the tunneling spectra using the tunneling conductance well above Δ, i.e. $G_N(V) = \frac{G(V)}{G(V=3.5\ meV)}$. $G_N(0)$ shows a very narrow distribution sharply peaked at 0.05 consistent with a fully gapped superconducting state (Fig. **3e**). On the other hand the coherence peak height ($G_{Np}$), extracted from the average of the maxima of $G_N(V)$ at positive and negative bias voltages, shows a large spatial variation. It has been demonstrated from Quantum Monte Carlo simulations that in a disordered superconductor $G_{Np}$ provides a measure of the local superconducting order parameter[31]. We observe that $G_{Np}$ varies smoothly from 1.1-1.8 (Fig **1b**) forming an inhomogeneous structure varying smoothly over tens of nanometers length-scale. However, the coherence peak is present at all locations showing that the superconducting order parameter is finite at all points.

We now turn focus on the temperature variation of Δ. To obtain Δ, we fit the $G_N(V)$ $vs.$ $V$ spectra averaged over the entire area (Fig. **1c**) with the tunneling equation, $G(V) \propto \int_{-\infty}^{\infty} N_s(E)\left(-\frac{\partial f(E-eV)}{\partial E}\right)dE$. Here *e* is the electronic charge, *f(E)* is the Fermi Dirac distribution function, and $N_s(E) = \frac{|E|+i\Gamma}{\sqrt{(|E|+i\Gamma)^2-\Delta^2}}$ is the BCS quasiparticle density of states, where a phenomenological broadening parameter Γ is incorporated to take into account non-thermal sources of broadening in the density of states[32]. Comparing with the temperature variation of resistance, we observe that Δ follows usual BCS temperature dependence and vanishes at $T_c$. Γ on the other hand is nearly temperature independent and varies between 0.23-0.33. Since the dominant role of Γ is to broaden the coherence peaks, the relatively large value of Γ reflects the variation in the coherence peak height.

*Emergence of granular superconducting state in magnetic field*

When a magnetic field is applied, the field enters a Type II superconductor in the form of vortices comprising of circulating supercurrent and enclosing a magnetic flux quantum, $\Phi_0$=*h*/2*e*. At the center of each vortex is the vortex core, where the superconducting order is destroyed and the circulating supercurrent is zero. In a STS experiment, these cores can be identified by recording the conductance map with bias voltage close to the coherence peak, where the vortices appear as low conductance points owing to the suppression of the coherence peak[33]. To identify the vortex cores in our sample, we record the conductance maps at 450 mK in different magnetic fields (H ⊥ film surface), at a fixed bias voltage of 2.2 mV (Fig. **2a-c**). The area was kept the same for different fields and small drifts ( < 10 nm ) were corrected by looking at topographic features of the surface. To avoid larger drifts during large temperature sweeps, the sample was first cooled in zero field and the data was taken by gradually ramping up the field and stabilizing at specific values. In principle, this zero field cooled protocol could suffer from the drawback that in a disordered superconductor the flux density at the center of the sample might be lower than the applied field owing to strong pinning. To verify whether the magnetic field is uniformly entering at the field where we performed the STS measurements, careful magnetization measurements performed on the field cooled and zero field cooled state as a function of temperature. We observed that the magnetization of the two states become indistinguishable above 5 kOe (see **Appendix A**), confirming that the flux completely penetrates the sample above this field. Identifying the vortices from the local minima in $G(V=2.2\ mV)$ we observe that at 40 kOe lines of vortices form a laminar structure, separating regions where $G(V=2.2\ mV)$ is high. As the field is increased, further entry of vortices progressively widen the regions with suppressed coherence peak and the laminar structure becomes denser, thereby shrinking the puddles where the coherence peak is high. We would like to note that at 60 and 75 kOe counting the number of vortices we observe that multiplying the number of vortices with $\Phi_0$ does not account for the entire flux expected to pass through the area for the applied magnetic field. This discrepancy arises because of our inability to account for two very closely placed vortices which appear as a single patch where the coherence peak is suppressed. This is consistent with numerical simulations that we present later.

We now investigate the nature of the vortex core, by recording the tunneling spectra along a line passing through the center of a vortex (Fig. **2d-f**). In a conventional superconductor, the core of an Abrikosov vortex behaves like a normal metal, where the tunneling spectrum is either flat ( $G_N(V) \sim 1$ ) or displays a small peak at zero bias owing to the formation of Caroli-De Gennes-Matricon[34] (CDM) bound states in very clean samples (see **Appendix B**). In contrast here, we observe that a soft-gap continues to survive even at the center of the vortex core (Fig. **2h-j**) although the coherence peak gets suppressed[35]. The suppression of the coherence peak suggests that the superconducting order-parameter is suppressed in the core of the vortex even though the pairing amplitude remains finite.

Since the proliferation of vortices locally suppresses the superconducting order, the superconducting state gets fragmented by lines of vortices into weakly coupled superconducting patches. To visualize this fragmented superconducting state more clearly, we measured the tunneling conductance spectra over the same area on a 32×32 grid in different magnetic fields. Fig. **3a-d** show the ZBC maps at different fields up to 75 kOe (corresponding to the same area as in Fig. **2a-c**). We observe that with increase in field the superconducting state develops large inhomogeneity, fragmenting into regions where $G_N(0)$ is large and regions where $G_N(0)$ is small. This is also reflected in the distribution of $G_N(0)$, which with increase in field develops large tails (Fig. **3e-h**). Fig. **3i-k** show the coherence peak height maps corresponding to the same fields. We observe an inverse correlation of the $G_{Np}$ maps with the $G_N(0)$ maps, implying that regions where $G_N(0)$ is large, the coherence peak is suppressed. The anticorrelation is also apparent from the 2D histogram of $G_N(0)$ and $G_N(V = 2.2$ mV$)$ which shows a negative slope over a large scatter, which suggests that the anti-correlation is not perfect. We quantify the anti-correlation using the cross-correlator, $I = \frac{1}{n}\sum_{i,j} \frac{\left(G_N^{i,j}(0) - \langle G_N(0)\rangle\right)\left(G_N^{i,j}(2.2\ mV) - \langle G_N(2.2\ mV)\rangle\right)}{\sigma_0 \sigma_{2.2\ mV}}$ , where $\sigma_0$ and $\sigma_{2.2\ mV}$ are the standard deviations in the values of $G_N(0)$ and $G_N(V = 2.2$ mV$)$ respectively, $i,j$ refer to the pixel index of the image and $n$ is the total number of pixels. We obtain $I \sim -0.15 - -0.2$ where $I = -1$ implies perfect anti-correlation. This is qualitatively consistent with earlier observation in strongly disordered NbN samples in zero field[7]. As expected, the vortices (shown in red rods) are preferentially located in the regions where $G_N(0)$ is high.

***The field induced pseudogap state***

We now investigate the temperature evolution of the superconducting state in magnetic field. Fig. **4a-e** show the temperature variation of the average $G_N(V)$-$V$ spectra over a 200 nm × 200 nm area at different magnetic fields along with the temperature variation of resistance. As the magnetic field is increased we observe that a soft gap in the tunneling spectrum continues to persist up to a temperature, $T^*$, well above $T_c(H)$. (For consistency, we define $T^*$ as the temperature where the zero bias conductance is 95% of the normal state value.) Here, $T_c(H)$ is defined as the temperature where the resistance is 0.05 % of its normal state value (for more details see **Appendix C**). This is analogous to the pseudogap state observed earlier in zero field in strongly disordered superconductors[4,5,7,15]. Plotting $T_c(H)$ and $T^*$ in the $H$-$T$ parameter space ( Fig. **4f** ), we observe that the pseudogap state becomes progressively wider as the magnetic field is increased. To rule out the possibility that the observed pseudogap is caused from a local distribution in temperature at which $\Delta \rightarrow 0$, we have also separately tracked the temperature dependence of the tunneling spectra at locations where the coherence peak at low temperature is finite and locations where the coherence peak is suppressed. Fig. **4h-i** show the temperature evolution corresponding to two such locations (Fig. **4g**) at 40 kOe. We observe that at both locations $G_N(0) \rightarrow 1$, at the same temperature, confirming that the pairing amplitude uniformly disappears at the same temperature (Fig. **4j**).

**IV. Comparison with numerical simulations**

We next carry out numerical simulation in order to develop further insight to our experimental findings. We describe our system through an attractive Hubbard Hamiltonian, which in the presence of disorder and applied magnetic field has the form,

$$H = -t \sum_{\langle i,j \rangle, \sigma} e^{i\varphi_{ij}} c_{i\sigma}^{+} c_{j\sigma} - |U| \sum_i \hat{n}_{i\uparrow} \hat{n}_{i\downarrow} + \sum_{i,\sigma} (V_i - \mu) \hat{n}_{i\sigma} , \qquad (1)$$

where $c_{i\sigma}$ ($c^+_{i\sigma}$) annihilates (creates) an electron with spin σ at site $i$ of a two dimensional square lattice, $\hat{n}_{i\sigma} = c^+_{i\sigma} c_{i\sigma}$ is the occupation number of site $i$ with spin σ and the phases $\varphi_{ij} = \frac{\pi}{\Phi_0} \int_i^j \overline{A} \cdot d\overline{l}$ are the Peierls factor of an applied orbital magnetic field. We use the Landau gauge $\overline{A} = Bx\hat{y}$ for all our calculations. The attraction $U$ induces s-wave superconductivity in the system. The disorder at site $i$ is given by $V_i$, which is chosen as independent random variable from a uniform distribution between $-V$ to $V$ which quantifies the disorder strength $V$. The chemical potential $\mu$ fixes the average density ($\rho = \sum_i \hat{n}_i$) which we fix at $\rho = 0.875$ for all our calculations. We carry out a fully self-consistent mean field analysis of (1) using Bogoliubov-de Gennes technique following ref. 36,37,38 on a 36 × 36 two-dimensional grid. We use $|U| = 1.2t$ which is much larger than in the weak-coupling BCS value. This is necessitated by the need to keep the coherence length within the system size. In the clean limit this gives a coherence length, $\xi_c$ ~ 10-12 lattice spacing, which is further reduced in the presence of disorder to give an operational coherence length, $\xi \sim (\xi_c l)^{0.5}$ ~ 5-6 lattice spacing[36]. For the same reason the effective magnetic field for finite number of vortices ($n$) is much larger than experimental value. Despite this caveat, it has been shown through several studies that the finite size simulations capture the broad qualitative features of disordered superconductors[1,2,24,31,36]. Since these simulations are restricted to $T = 0$, we compare our simulations with experimental data taken at the lowest temperature, 450 mK.

We first investigate the zero field state obtained from the simulations. Fig. **5a** shows the single particle density of states, $D(E)$, normalized to its value at $E = -0.2t$, averaged over the entire lattice for $V = 0.5t$. The average $D(E)$ shows a fully formed gap and sharp coherence peaks consistent with the zero field tunneling spectra at 450 mK. Fig. **5b** shows the spatial variation of $D(E)$ at the coherence peak, $D_p$. We observe that $D_p$ shows large spatial variation forming an inhomogeneous structure similar to that in Fig. **1b**. To confirm that this disorder strength $V$ is indeed appropriate for our experiments, we compare the width of the normalized distribution of $G_{Np}$ at 450 mK in zero field, defined as $\tilde{G}_{Np} = \frac{G_{Np} - G_{Np}^{Min}}{G_{Np}^{Max} - G_{Np}^{Min}}$ with the corresponding normalized distribution of $D_p$, namely, $\tilde{D}_p = \frac{D_p - D_p^{Min}}{D_p^{Max} - D_p^{Min}}$ (where $G_{Np}^{Min}(D_p^{Min})$ and $G_{Np}^{Max}(D_p^{Max})$ are the minimum and maximum value of $G_{Np}(D_p)$). The distributions (Fig. **5c**) have similar width as measured from standard deviations from the mean value showing that we are working at comparable disorder strength.

We now track the evolution of the superconducting state with magnetic field. Since the vortex simulations are carried out by repeating the unit cells periodically, $n$ can only be even[37]. Fig. **6a-6c** show the spatial variation of the phase of the superconducting order parameter, $\phi$, for $n = 2, 4$ and $6$; the color scale corresponds to the local pairing amplitude defined as, $|\Psi_i| = |U||\langle c_{i\downarrow} c_{i\uparrow}\rangle|$. The positions of the vortices can be identified from the locations where $\phi$ twists around a point and $|\Psi| \sim 0$. The spatial variation of $D(0)$ and $D_p$ corresponding to these flux filling shown in Fig. **6d-6i** qualitatively capture all the broad features observed in our experiment. The presence of the vortex results in a local increase in $D(0)$ and a concomitant suppression in $D_p$. (The anti-correlation between $D(0)$ and $D_p$ is less clear for $n = 6$, possibly due to the large effective field in our simulations which suppresses the coherence peaks). Furthermore, we observe from the $D(0)$ ($D_p$) maps that the regions with large (small) $D(0)$ ($D_p$) around two closely located vortices coalesce to form one continuous larger patch. (The length-scale of these patches is of the order of the coherence length.) Therefore, it is likely that in our in-field experiments we are unable to resolve all the individual vortices from the conductance images, which accounts for the apparent non-conservation of magnetic flux in our sample. We also observe that the distribution of $D(0)$ progressively increases with increasing $n$ (Fig. **6j-6l**) and forms long tails consistent with experiments.

Finally, we dwell on the issue of soft gap observed inside the vortex core in our experiments. In Fig. **6k** (inset) we compare the average $D(E)$ close to the center of the vortices and at regions far from for $n = 4$. Close to the center of the vortices the coherence peak is completely suppressed but a soft gap continues to survive. For consistency check we have performed the same calculations without any disorder ($V=0$) (see **Appendix D**). In that case, we realize an Abrikosov lattice commensurate with the lattice geometry, and $D(E)$ shows a large zero energy peak inside the vortex core consistent with the CDM bound state[34]. To understand physically the origin of this behavior we note that the circulating supercurrent density around a vortex, $\boldsymbol{J} \propto (1/r)$, where $r$ is the radial distance from the center of the vortex. In a clean superconductor, the normal vortex core appears below a limiting $r$, where the increase in kinetic energy of the Cooper pair exceeds the pairing energy, $2\Delta$, and destroys the superconducting pairing. In the presence of disorder a completely different scenario can emerge. Here, disorder scattering reduces the superfluid stiffness, $J_s$, making the superconductor susceptible to phase fluctuations[28]. The survival of the soft gap in the vortex core, in our opinion, is strongly tied to the phase fluctuations of the order parameter due to the inhomogeneous background that depletes the superfluid stiffness in the core regions, but keeps the pairing amplitude finite.

**V. Summary and Outlook**

The emerging physical picture from the experiments and the simulations is as follows: The random disorder potential makes superconducting order parameter spatially inhomogeneous even in the absence of magnetic field. As a result, the flux tubes from applied field threads the system through locations where the local amplitude of the order parameter, |Ψ|, is low. Such spatial organization lessens the energy cost by accumulating the phase-twist in regions of low |Ψ|. This naturally makes the field induced vortex lattice aperiodic, and flux tubes wipe out remnants of pairing amplitude in a region of size ~ ξ around vortex centers. Such local annihilation of superconducting correlations with magnetic field introduces granularity in the superconducting state even with low disorder strengths, in a manner similar to what disorder alone would have done for much larger strengths. Consequently, the pseudogap is observed in the region of *H-T* parameter space where Cooper pairs continue to survive even when the zero resistance state is destroyed due to phase fluctuation between superconducting puddles. This is consistent with earlier planar tunneling measurements on Pb-Bi films[39].

Our results provide valuable clue to understand the magnetic field induced superconductor-insulator transition in much more strongly disordered samples, which are very close to, but on the superconducting side of disorder driven SIT. There, even the zero field state consists of regions where |Ψ| is completely suppressed such that the superconducting state is composed of superconducting puddles that are Josephson coupled through insulating regions. When a magnetic field is applied, the superconducting puddles will further fragment through vortex proliferation, till they reach a critical size where Coulomb blockade makes it energetically unfavorable for the current to pass through the superconducting islands[24]. At this point we would expect to see a transition from a superconductor to an insulator-like behavior in transport measurements. Therefore, we propose that the disorder and magnetic field driven SIT-s are both manifestation of the same microscopic phenomenon: The granularity that emerges naturally in the superconducting state. Microscopic validation of this scenario could be obtained through STS measurements on more strongly disordered superconductors at very low temperatures.

## Appendix A

*Flux penetration in the superconductor in the zero field cooled state*

In a strongly pinned superconductor, when magnetic field is applied after the superconductor is cooled to low temperatures in zero magnetic field (i.e. the zero field cooled (ZFC) state), the entry of magnetic flux is hindered by the pinning potential. Consequently, the flux density gradually decays from the edge towards the center of the superconductor, where the flux density gradient is determined by the local critical current density[40]. In contrast, when the sample is cooled from above $T_c$ in the presence of magnetic field (i.e. the field cooled (FC) state) the presence of strong pinning traps the magnetic flux threading the sample in the normal state in the form of vortices, producing a nearly uniform flux density profile, and a magnetization (**M**) that is higher than in the ZFC state. As the magnetic field is gradually increased towards larger values, the critical current density of the superconductor decreases and the difference between the FC and ZFC state becomes smaller.

To assess the flux penetration in our sample we performed careful **M**-*T* measurements in the ZFC and FC state for different magnetic fields using SQUID magnetometer (Fig. **7**). The ZFC state is created by applying the magnetic field after cooling the sample to the base temperature (1.8 K) in zero field. The FC state is created by applying the field at 15 K and cooling the sample to the base temperature in magnetic field. The **M**-*T* measurements are carried while warming up the sample from this initial ZFC or FC state. At 10 Oe the ZFC curve shows pronounced diamagnetic response whereas the FC curve is flat and very close to zero as expected for a strongly pinned Type II superconductor. However, as magnetic field is increased the difference between the FC and ZFC curves progressively decreases, and at 5 kOe the two become indistinguishable within experimental resolution. Thus beyond this field the flux completely enters the ZFC state and the role of flux pinning on the entry of flux in the ZFC state is negligible.

## Appendix B

*Comparison between Abrikosov vortex cores and the vortex core in disordered NbN*

Here we compare the vortex core in disordered NbN with the Abrikosov vortex core in a clean single crystal of NbSe$_2$ ( $T_c$ ~ 7.2 K ). Fig. **8a** and Fig. **8d** show the images of the vortices acquired at 450 mK in NbN and in a pure NbSe$_2$ crystal respectively. Fig. **8b** and Fig. **8e** show the normalized tunneling spectra along a line passing through the center of the vortex for the two samples. For NbSe$_2$ we observe that for the normalized tunneling spectrum at the center of the vortex G(V) ≥ 1 at all bias and the spectrum shows a zero bias conductance peak associated with Caroli-de Gennes-Matricon (CDM) state[34,41] (Fig. **8f**). In contrast, at the center, the normalized tunneling spectrum for NbN shows a pseudogap (Fig. **8c**), characterized by a suppression of the coherence peak and a soft-gap characteristic of superconducting pairing.

## Appendix C

*Criterion for the upper critical field from transport measurements*

While determining $H_{c2}(T)$ (or equivalently, $T_c(H)$) from transport measurements different criteria are used in the literature. While sometimes it is defined as the locus of points where the resistance drops to 90% of its normal state value, in other instances it is defined as the locus of points where the resistance falls below the measurable limit.

To determine the most appropriate criterion in our context, we measured the diamagnetic shielding response of the superconducting film using a two coil mutual inductance technique[42,43]. In this technique the superconducting film is sandwiched between a quadrupolar primary coil and a dipolar secondary coil and the mutual inductance ($m$) is measured between the two (Fig **9a**). Below the superconducting transition, the superconducting film partially shields the magnetic field produced by the primary coil from the secondary coil and real part of the mutual inductance ($m'$) decreases, signaling the onset of the diamagnetic response.

Fig. **9b** shows the $m'$-$H$ (upper panel) and $R$-$H$ (lower panel) measured at different temperatures. To measure $m'$ we use an a.c. excitation field with amplitude 3.5 mOe and frequency of 31 kHz. We observe that the onset of a.c. shielding response coincides with the field where the resistance drops to 0.05 % of its normal state value, below which the resistance measurement goes below our lower measurable limit. We define this field as our upper critical field[44]. We conclude that the broad transition region above this field observed in $R$-$H$ measurements consists of phase fluctuating superconducting puddle where the global superconducting order is destroyed. The same locus of points in the $H$-$T$ parameter space can also be obtained by measuring $T_c(H)$ from $R$-$T$ measurements in constant magnetic fields, where $T_c(H)$ defined as the temperature where the resistance drops to 0.05 % of its normal state value (Fig. **9c**).

## Appendix D

*Details of numerical simulations*

We start our model calculation by benchmarking the numerical simulation for a clean ( $V = 0$ ) s-wave superconductor in orbital field. We work with Landau gauge, so that the vector potential for desired field $B\hat{z}$ becomes $\mathbf{A} = (0, Bx)$. For our BdG solution we always consider a rectangular unit cell of size $L_x \times 2L_x$ that contains two superconducting flux quanta. We tune the magnetic field strength $B$ by changing the system size $L_x$. Our construction thus forces a square vortex lattice. This commensurability of the vortex lattice with the underlying lattice on which the electrons live is unavoidable for a simulation box that is not too large, like in the present case. The translation symmetry of the vortex lattice by multiples of inter-vortex distances allows the centre of the two flux tubes to appear anywhere in the system provided their relative distance is $L_x$, and we choose them to lie at the centre of each square-shaped half unit-cell. We start the BdG calculation with a guess profile of $\Delta(r)$ taken as the analytical solution of Abrikosov close to $H_{c2}$ [45]. This allows a fairly good starting point for the phase of the order parameter, aiding convergence of the BdG self-consistency significantly. The vortex lattice is generated in simulation taking advantage of Bloch-translation in which the BdG solution in a unit-cell is periodically repeated[37]. The procedure outlined here reproduces standard results, e.g. the depletion of pairing amplitude at the vortex core of diameter ~ $\xi$, expected curling of the phase of the order parameter, and the zero-bias peak at $E/t= 0$ in the local density of states at the vortex core due to the formation of the CDM bound state[34]. In the clean system we work with a 40 × 80 rectangular lattice. Here we only show the square area through which only one SC flux passes. Such structures are repeated periodically. These results are shown in Fig. **10a-c**.

Upon implementing the simulation of a clean vortex lattice that reproduces standard results, we proceed to include disorder. Lack of translation symmetry in the presence of disorder does not offer a good guess for the local order parameter causing the convergence to self-consistency very slow. We used combinations of Anderson, Broyden and Modified Broyden mixing methods[38] to accelerate the convergence to self-consistency. Even then, the number of iteration for self-consistency for $n = 2$ on a given realization of disorder at $V = 0.5$ grows by two orders of magnitude compared to $V = 0$ case. As a consequence we cannot simulate large unit cell unlike the clean case.

We work with square cell of size 36 × 36 and increase the field by changing the number of flux through a unit cell for the disordered case. Confidence in our results was developed by choosing different initial conditions and arriving at the same final self-consistent value of the complex order parameter.

**Author Contribution**

RG and IR performed the measurements. IR and RG analyzed the data. HS prepared the sample. AB carried out the simulations under the supervision of AG. PR conceived the problem and supervised the experiments. PR and AG wrote the paper. All authors discussed the results and commented on the manuscript. RG and IR contributed equally as joint first author of this paper.

**Acknowledgements:** We would like to thank Jim Valles, Nandini Trivedi, Lara Benfatto and Vikram Tripathi for useful discussions. The work was supported by Department of Atomic Energy, Govt. of India and Department of Science and Technology, Govt of India (Grant No: EMR/2015/000083).

[1] A. Ghosal, M. Randeria & N. Trivedi, Role of Spatial Amplitude Fluctuations in Highly Disordered s-wave Superconductors. Phys. Rev. Lett. 81, 3940–3943 (1998).

[2] Y. Dubi, Y. Meir & Y. Avishai, Nature of the superconductor-insulator transition in disordered superconductors. Nature 449, 876–880 (2007).

[3] M. V. Feigel'man, L. B. Ioffe, V. E. Kravtsov & E. A, Yuzbashyan, Eigenfunction Fractality and Pseudogap State near the Superconductor-Insulator Transition. Phys. Rev. Lett. 98, 027001 (2007).

[4] B. Sacépé et al, Disorder-induced inhomogeneities of the superconducting state close to the superconductor-insulator transition. Phys. Rev. Lett. 101, 157006 (2008).

[5] M. Mondal et al, Phase Fluctuations in a Strongly Disordered s-Wave NbN Superconductor Close to the Metal-Insulator Transition. Phys. Rev. Lett. 106, 047001 (2011).

[6] B. Sacépé et al, Localization of preformed Cooper pairs in disordered superconductors. Nat. Phys. 7, 239–244 (2011).

[7] A. Kamlapure et al, Emergence of nanoscale inhomogeneity in the superconducting state of a homogeneously disordered conventional superconductor. Sci. Rep. 3, 2979 (2013).

[8] G. Lemarie et al, Universal scaling of the order-parameter distribution in strongly disordered superconductors. Phys. Rev. B 87, 184509 (2013).

[9] D. Sherman et al, The Higgs mode in disordered superconductors close to a quantum phase transition, Nat. Phys. 11, 188–192 (2015).

[10] A Kamlapure, S Manna, L Cornils, T Hänke, M Bremholm, Ph Hofmann, J Wiebe, R Wiesendanger, Spatial variation of the two-fold anisotropic superconducting gap in a monolayer of $FeSe_{0.5}Te_{0.5}$ on a topological insulator. Phys. Rev. B 95, 104509 (2017).

[11] M. Tinkham, Introduction to Superconductivity. (Dover Publications Inc., Mineola, New York, 2004).

[12] P. W. Anderson, Theory of Dirty Superconductors. J. Phys. Chem. Solids 11, 26–30 (1959).

[13] M. Ma & P. Lee, Localized superconductors. Phys. Rev. B 32, 5658–5667 (1985).

[14] G. Seibold, L. Benfatto, C. Castellani & J. Lorenzana, Superfluid Density and Phase Relaxation in Superconductors with Strong Disorder. Phys. Rev. Lett. 108, 207004 (2012).

[15] B. Sacépé et al, Pseudogap in a thin film of a conventional superconductor. Nat. Commun. 1, 140 (2010).

[16] C. Carbillet et al., Confinement of superconducting fluctuations due to emergent electronic inhomogeneities. Phys. Rev. B **93**, 144509 (2016).

[17] D. Sherman, G. Kopnov, D. Shahar & A. Frydman, Measurement of a Superconducting Energy Gap in a Homogeneously Amorphous Insulator. Phys. Rev. Lett. 108, 177006 (2012).

[18] T. I. Baturina et al, Localized Superconductivity in the Quantum-Critical Region of the Disorder-Driven Superconductor-Insulator Transition in TiN Thin Films. Phys. Rev. Lett. 99, 257003 (2007).

[19] G. Sambandamurthy, L. W. Engel, A. Johansson & D. Shahar, Superconductivity-Related Insulating Behavior. Phys. Rev. Lett. 92, 107005 (2004).

[20] M. D. Stewart, Jr., A. Yin, J. M. Xu & J. M. Valles, Jr., Magnetic-field-tuned superconductor-to-insulator transitions in amorphous Bi films with nanoscale hexagonal arrays of holes. Phys. Rev. B 77, 140501(R) (2008).

[21] M. D. Stewart, Jr., A. Yin, J. M. Xu & J. M. Valles, Jr., Superconducting Pair Correlations in an Amorphous Insulating Nanohoneycomb Film. Science 318, 1273-1275 (2007).

[22] G. Kopnov, O. Cohen, M. Ovadia, K. Hong Lee, C. C. Wong & D. Shahar, Little-Parks Oscillations in an Insulator. Phys. Rev. Lett. 109, 167002 (2012).

[23] S. M. Hollen, G. E. Fernandes, J. M. Xu & J. M. Valles Jr, Collapse of the Cooper pair phase coherence length at a superconductor-to-insulator transition. Phys. Rev. B, **87**, 054512 (2013).

[24] Y. Dubi, Y. Meir & Y. Avishai, Theory of the magnetoresistance of disordered superconducting films. Phys. Rev. B 73, 054509 (2006).

[25] A. Gangopadhyay, V. Galitski & M. Müller, Magnetoresistance of an Anderson Insulator of Bosons. Phys. Rev. Lett. 111, 026801 (2013).

[26] V. M. Vinokur et al, Superinsulator and quantum synchronization, Nature 452, 613-615 (2008).

[27] M. Ovadia, D. Kalok, B. Sacépé & D. Shahar, Duality symmetry and its breakdown in the vicinity of the superconductor–insulator transition. Nat. Phys. 9, 415–418 (2013).

[28] M. Chand et al, Phase diagram of a strongly disordered s-wave superconductor, NbN, close to the metal-insulator transition. Phys. Rev. B 85, 014508 (2012).

[29] "Weak disorder" in the present paper refers to the Regime **I** described in ref. 28.

[30] A. Kamlapure et al, A 350 mK, 9 T scanning tunneling microscope for the study of superconducting thin films on insulating substrates and single crystals. Rev. Sci. Instr. 84, 123905 (2013).

[31] K. Bouadim, Y. L. Loh, M. Randeria & N. Trivedi, Single- and two-particle energy gaps across the disorder-driven superconductor–insulator transition. Nature Physics 7, 884–889 (2011).

[32] R. C. Dynes, V. Narayanamurti & J. P. Garno, Direct Measurement of Quasiparticle-Lifetime Broadening in a Strong-Coupled Superconductor. Phys. Rev. Lett. 41, 1509 (1978).

[33] S. C. Ganguli et al, Disordering of the vortex lattice through successive destruction of positional and orientational order in a weakly pinned $Co_{0.0075}NbSe_2$ single crystal. Scientific Reports 5, 10613 (2015).

[34] C. Caroli, P. G. De Gennes & J. Matricon, Bound Fermion states on a vortex line in a type II superconductor. Physics Letters, Volume 9, 307-309 (1964).

[35] Y. Noat et al, Unconventional superconductivity in ultrathin superconducting NbN films studied by scanning tunneling spectroscopy. Phys. Rev. B 88 014503 (2013).

[36] A. Ghosal, M. Randeria & N. Trivedi, Inhomogeneous pairing in highly disordered s-wave superconductors. Phys. Rev. B **65**, 014501 (2001).

[37] A. Ghosal, C. Kallin & A. J. Berlinsky, Competition of superconductivity and antiferromagnetism in a d-wave vortex lattice. Phys. Rev. B 66, 214502 (2002).

[38] D. D. Johnson, Modified Broyden's method for accelerating convergence in self-consistent calculations. Phys. Rev. B 38, 12807 (1988).

[39] S. Y. Hsu, J. A. Chervenak & J. M. Valles Jr, Magnetic field enhanced order parameter amplitude fluctuations in ultrathin films near the superconductor-insulator transition. Phys. Rev. Lett. **75**, 132 (1995).

[40] C.P. Bean, Magnetization of High-Field Superconductors. Rev. Mod. Phys. 36, 31 (1964).

[41] H. F. Hess, R. B. Robinson, R. C. Dynes, J. M. Valles Jr & J. V. Waszczak, Scanning-tunneling-microscope observation of the Abrikosov flux lattice and the density of states near and inside a fluxoid. Phys. Rev. Lett. **62**, 214 (1989).

[42] S. Kumar, C. Kumar, J. Jesudasan, V. Bagwe, P. Raychaudhuri and S. Bose, A two-coil mutual inductance technique to study matching effect in disordered NbN thin films. Appl. Phys. Lett. **103**, 262601 (2013).

[43] I. Roy, P. Chauhan, H. Singh, S. Kumar, J. Jesudasan, P. Parab, R. Sensarma, S. Bose and P. Raychaudhuri, Dynamic transition from Mott-like to metal-like state of the vortex lattice in a superconducting film with a periodic array of holes. Phys. Rev. B **95,** 054513 (2017).

[44] This definition differs from the 90% criterion used is ref.28, where the $H_{c2}$ was determined from transport measurements alone.

[45] A. A. Abrikosov, On the Magnetic Properties of Superconductors of the Second Group. Sov. Phys. JETP **5,** 1174 (1957).

**Figure Captions**

**Figure 1| Superconducting state in zero field.** (a) Representative normalized tunneling conductance spectra ($G_N(V)$ vs. V) in zero field at 450 mK along a 200 nm line. The spectra show uniform energy gap and finite coherence peak at the gap edge over the entire line. (b) Coherence peak height ($G_{Np}$) map over a 200 nm x 200 nm area, forming an inhomogeneous structure. (c) Temperature variation of the average superconducting energy gap, $\Delta$, (green squares) and broadening parameter, $\Gamma$ (red circles) obtained by fitting the average $G_N$ (V) vs. V spectra. The fits to the tunneling spectra are shown in the inset. The light green line is the expected variation of $\Delta(T)$ obtained from BCS theory. The blue line shows temperature variation of resistance measured on the same sample. The resistance appears exactly at the same temperature, $T_c$, where BCS gap vanishes.

**Figure 2| Superconducting state in magnetic field.** (a)-(c) Conductance maps at 40 kOe, 60 kOe and 75 kOe respectively, taken at 450 mK at fixed bias 2.2 mV over an area of 200 nm x 200 nm. The red dots show the position of the vortices obtained from the local minima in the conductance. (d)-(f) Normalized tunneling spectra along the green lines shown in the conductance maps ((a)-(c)) respectively, which passes through the centre of the vortices. The vertical dashed line denotes the center of the vortex. (h)-(j) Representative spectra for three different fields respectively, at the center of the vortex core (black) and away from the core (violet). In contrast to a conventional superconductor, we observe a soft gap at the core of the vortices.

**Figure 3| Magnetic field induced granularity.** (a)-(d) ZBC ($G_N(0)$) maps for fields 0, 40, 60 and 75 kOe respectively over the same 200 nm x 200 nm area at 450 mK, obtained from area spectroscopy over 32x32 pixels grid. The red dots show the position of the vortices. (e)-(h) Distribution of $G_N(0)$ for 0, 40, 60 and 75 kOe respectively. With increasing field the distributions

develop large tails, signifying emerging inhomogeneity with field. (i)-(k) Coherence peak height ($G_{Np}$) maps for 40, 60 and 75 kOe. (l)-(n) Cross-correlation histograms between $G_N(0)$ and $G_{Np}$ for corresponding fields, showing inverse correlation between the two quantities.

**Figure 4| Field induced pseudogapped state.** (a)-(e) Temperature variation of the average $G_N(V)$-$V$ spectra at 0, 20, 40, 60 and 75 kOe respectively along with the temperature variation of resistance. The vertical dashed lines correspond to, $T_c$, where resistance appears and $T^*$, where the pseudogap in the density of states disappear. At $H=0$ these two happen at the same temperature. The range of the temperature axes in all plots has been kept same for visual comparison. (f) $T_c$ and $T^*$ are plotted on the $H$-$T$ space, which shows that the pseudogap state widens as the field is increased. (g) Conductance map at fixed bias, $V = 2.2$ mV at 40 kOe at 450 mK. The blue and green boxes show two representative areas where the conductance is high and low respectively. (h)-(i) Average $G_N(V)$-$V$ for different temperatures inside the blue and the green boxes respectively. (j) Temperature dependence of $G_N(0)$ for the average spectra inside the blue and green box respectively; we observe that in both cases $G_N(0)$ goes to 1 at the same temperature.

**Figure 5| Simulation of the superconducting state in zero field.** (a) Density of states, $D(E)$, in the absence of vortices averaged over a 36 × 36 lattice showing a fully formed gap and sharp coherence peaks. (b) Spatial variation of $D_p$ calculated for 36 × 36 lattice. (c) Normalized distribution of $\tilde{G}_{Np}$ obtained from experiments at 450 mK in zero field, and $\tilde{D}_p$. The distribution of $\tilde{G}_{Np}$ has a standard deviation, $\sigma_{\tilde{G}_{Np}} = 0.1358$, whereas the distribution of $\tilde{D}_p$ has a standard deviation, $\sigma_{\tilde{D}_p} = 0.141$.

**Figure 6| Simulation in the presence of vortices.** (a)-(c) The spatial variation of $\varphi$ shown as arrows for $n = 2$, 4 and 6 vortices on the 36×36 lattice. The colors of the arrows stand for the

strength of superconducting order parameter $|\Psi|$ on a scale of 0 to 1 on the lattice. The positions of the vortices can be identified from the locations where $\varphi$ twists around a point and $|\Psi|$ has a value close to zero. (d)-(f) The spatial variation of $D(0)$ corresponding to $n = 2$, 4 and 6 vortices on the 36×36 lattice. (g)-(i) The spatial variation of $D_p$ corresponding to $n = 2$, 4 and 6 vortices on the 36×36 lattice. (j)-(l) Distribution of $D(0)$ for $n = 2$, 4 and 6 vortices respectively. With increasing field the distributions develop large tails, signifying emerging inhomogeneity with field. The *inset of* (k) shows the *D(E)* as a function of *E/t* for $n = 4$ averaged over regions close to the center of the vortices (black) and for regions far from the vortices (blue).

**Figure 7| Flux penetration in the NbN film.** Temperature dependence of magnetization of the NbN film measured using the FC and ZFC protocol. The magnetization in the FC state is close to zero showing a nearly complete flux penetration. Above 3 kOe the magnetization of the FC and ZFC state are identical within experimental resolution. The substrate contribution in the magnetization has been subtracted from all curves.

**Figure 8| Comparison of vortex core in NbSe$_2$ and NbN.** (a) Conductance map over 200 nm x 200 nm area of NbN at 40 kOe at fixed bias voltage ($G(V=2.2\ mV)$). The vortices are shown as red dots. (b) Normalized tunneling spectra ($G_N(V)$-V) along the green line on conductance map in (a) going through the center of one vortex core; the dark black line corresponds to the spectrum at the vortex core. (c) Representative spectra for NbN at the center of the vortex core (black) and a point away from the core (violet). (d) Conductance map over 250 nm x 85 nm area of pure NbSe$_2$ at 7 kOe at fixed bias voltage ($G(V=1.3\ mV)$). The dark regions are the vortices with lower values of conductance forming a hexagonal Abrikosov lattice. (e) Tunneling spectra *($G_N(V)$-V)* along the green line on conductance map in (d) going through one vortex core of NbSe$_2$; dark black line

corresponds to the spectrum at the vortex core. (f) Representative spectra for NbSe$_2$ at the center of the vortex core (black) and a point away from the core (violet).

**Figure 9| Comparison of magnetic field variation of resistance and diamagnetic shielding response.** (a) Schematic diagram of the two-coil mutual inductance setup; the superconducting film is sandwiched between a quadrupolar primary coil and a dipolar secondary coil. (b) Magnetic field variation of m' (upper panel) and resistance (lower panel) at different magnetic fields; the vertical dashed lines show the onset of the diamagnetic shielding response which coincide with the field where the resistance goes below our measurable limit (at 0.05% of its normal state value). The resistance is plotted in log-scale for clarity. (c) The loci of $H_{c2}(T)/T_c(H)$ in the *H-T* parameter space from m'-*H*, *R-H* and *R-T* measurements.

**Figure 10| Vortex simulation in a clean superconductor.** (a) Surface plot of the pairing amplitude |Ψ| on a scale of 0 to 1 in a square unit cell illustrating the vortex. (b) Phase of the superconducting order parameter on a unit cell. (c) Local density of states for the vortex core region showing the CDM peak (black) and at regions away from the vortex core region (blue).

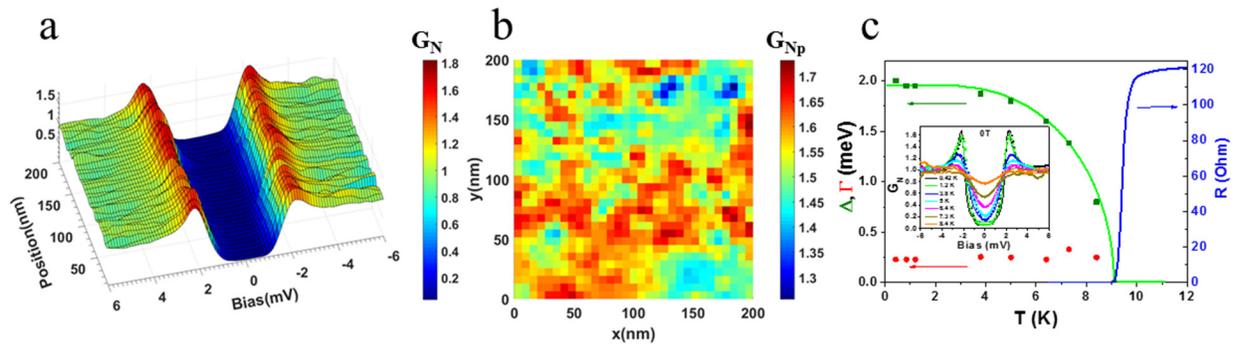

**Figure 1**

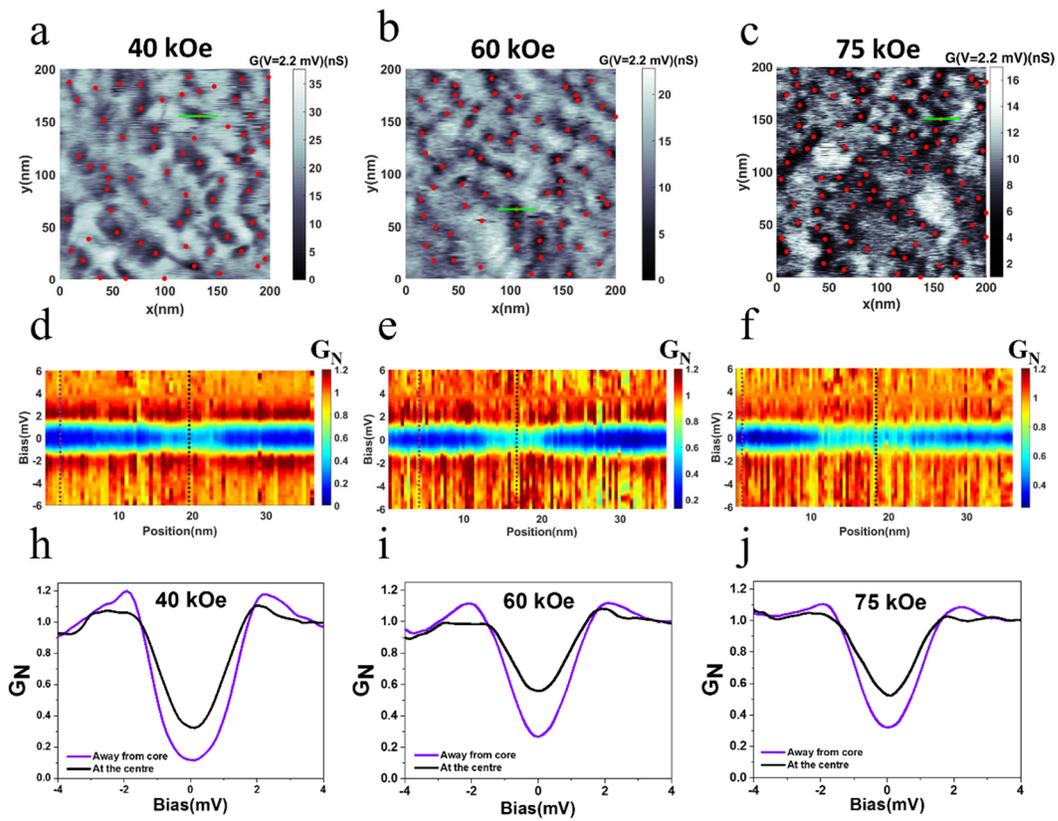

**Figure 2**

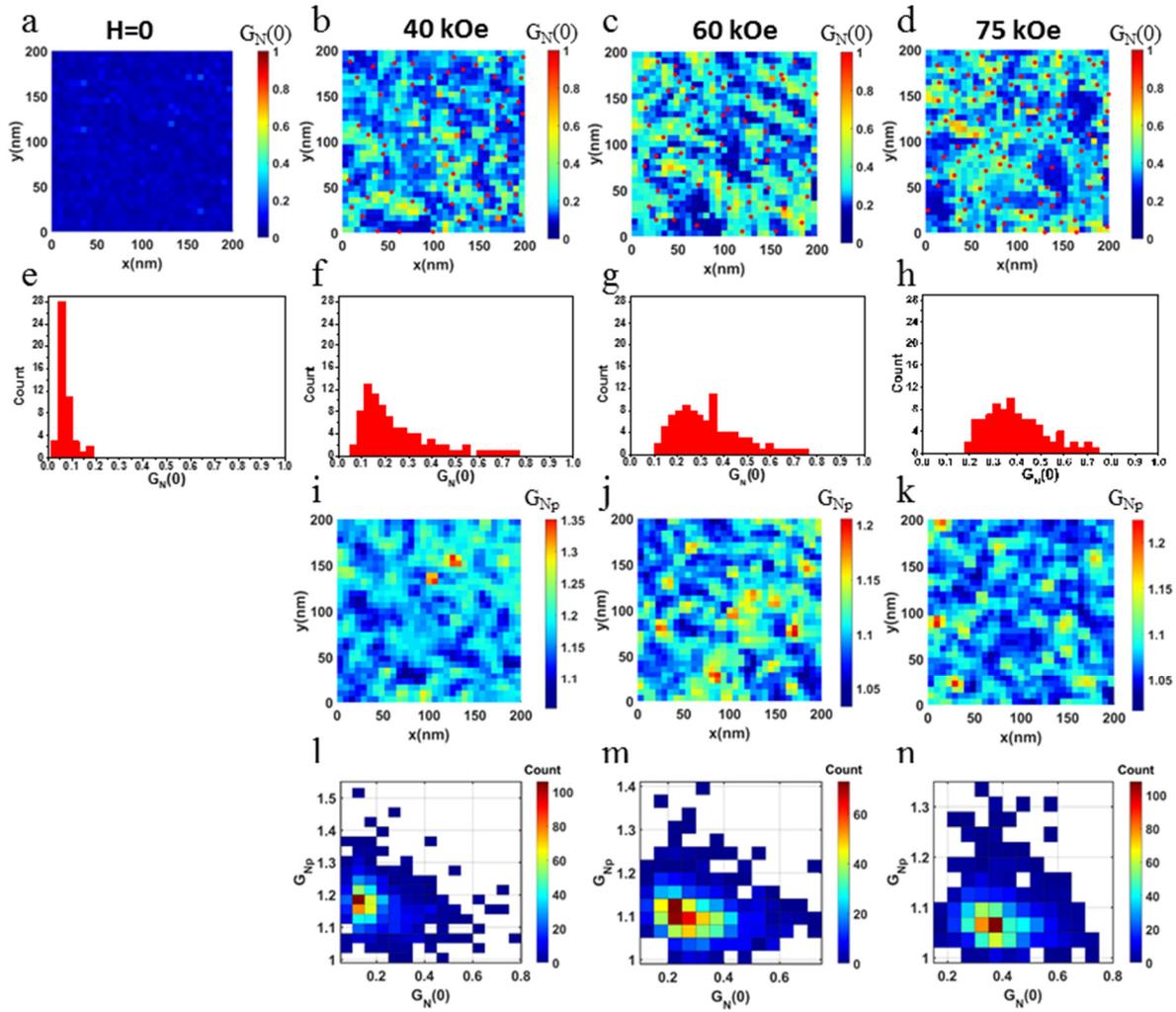

**Figure 3**

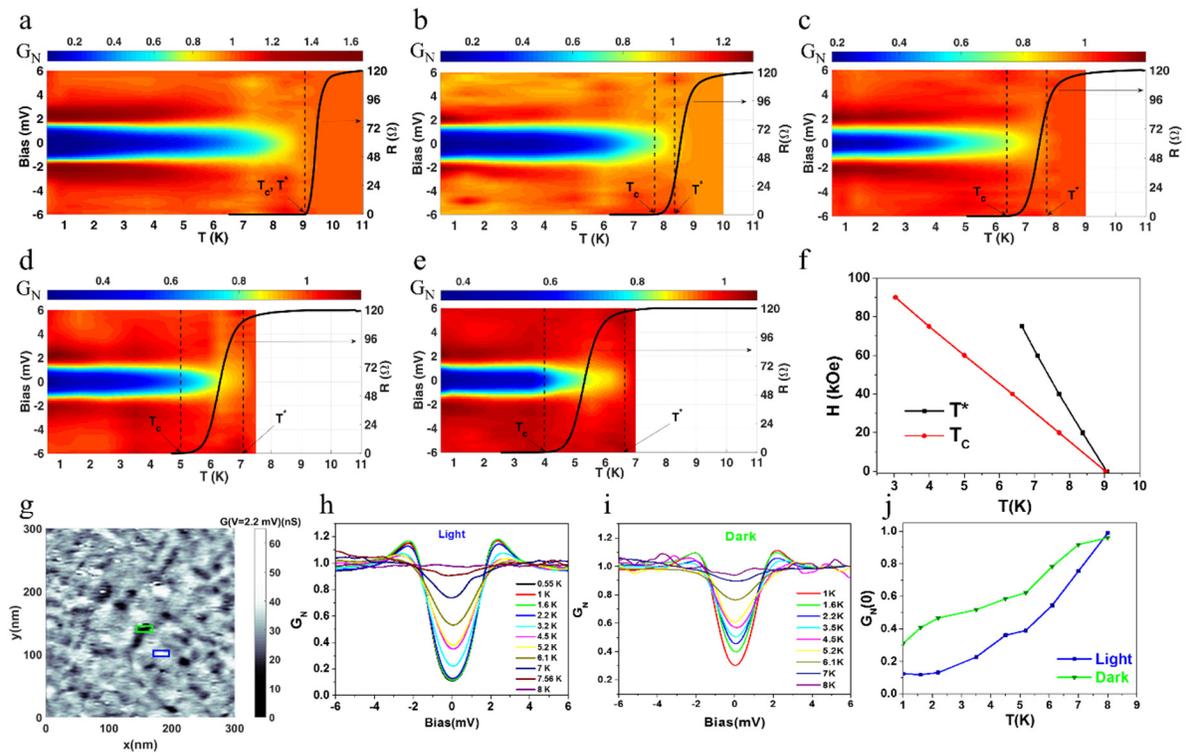

**Figure 4**

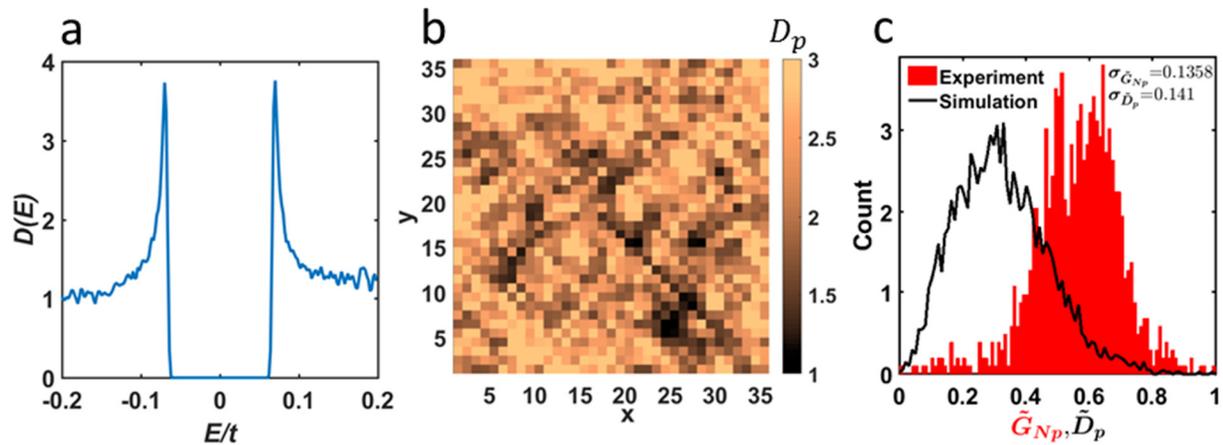

**Figure 5**

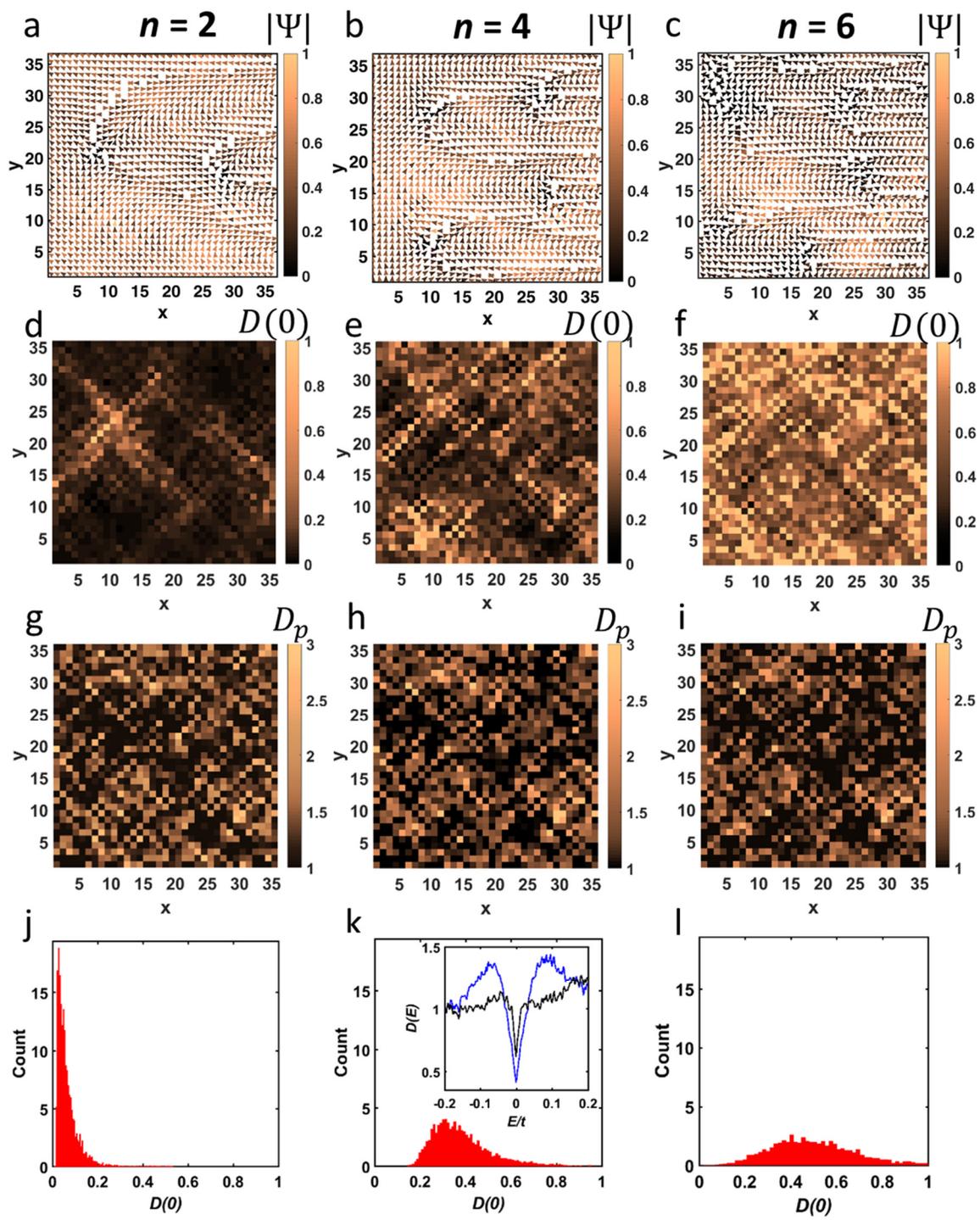

**Figure 6**

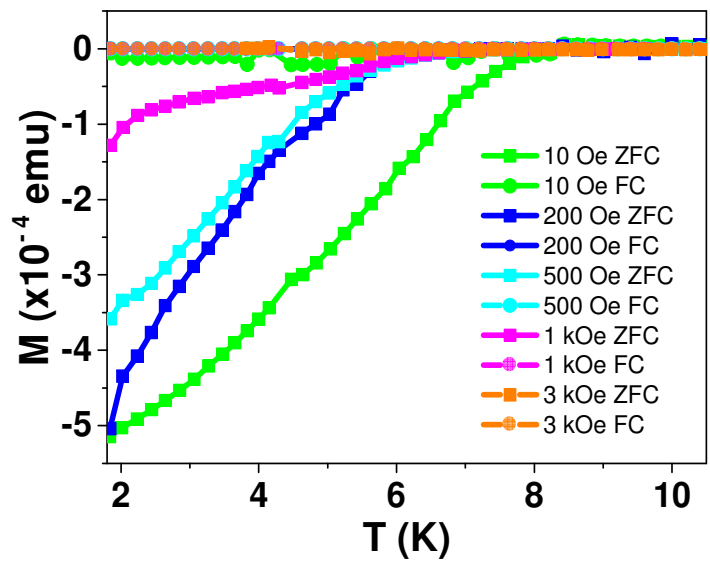

**Figure 7**

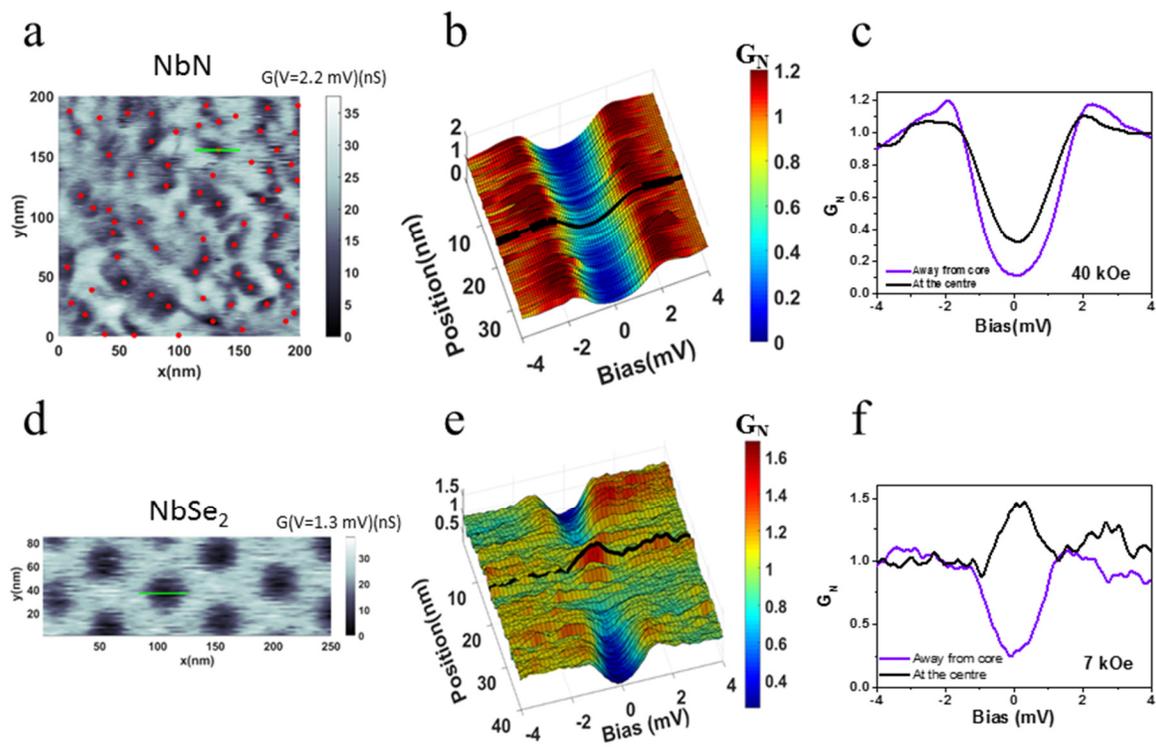

**Figure 8**

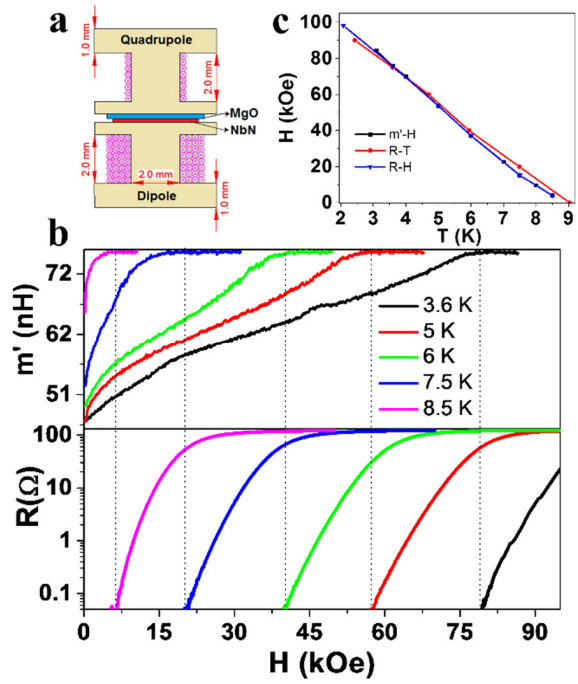

**Figure 9**

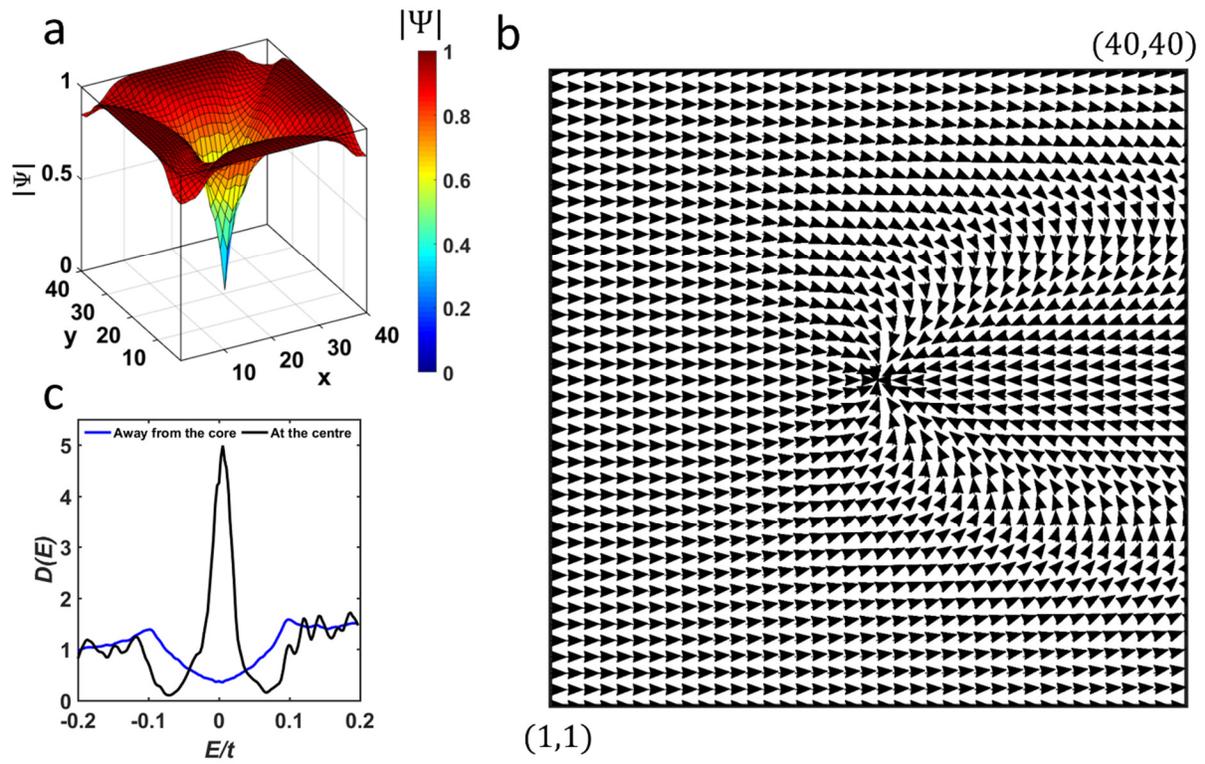

**Figure 10**